\documentclass[%
superscriptaddress,showkeys,
12pt,
amsmath,amssymb,
aps,
]{revtex4-1}

\usepackage{graphicx}
\usepackage{dcolumn}
\usepackage{bm}
\usepackage{xcolor}
\usepackage{url}
\usepackage{makecell}
\usepackage{subfigure}
\usepackage{multirow}
\usepackage{booktabs}
\usepackage{verbatim}
\usepackage{diagbox}
\usepackage[marginal]{footmisc}
\usepackage{amssymb}

\newcommand{\moe}{\affiliation{Key Laboratory of Atomic and Subatomic Structure and Quantum Control (MOE), Guangdong Basic Research Center of Excellence for Structure and Fundamental Interactions of Matter, Institute of Quantum Matter, South China Normal University, Guangzhou 510006, China
}}

\newcommand{\iqm}{\affiliation{Guangdong-Hong Kong Joint Laboratory of Quantum Matter, Guangdong Provincial Key Laboratory of Nuclear Science, Southern Nuclear Science Computing Center, South China Normal University, Guangzhou 510006, China}}

\newcommand{\scnt}{\affiliation{Southern Center for Nuclear-Science Theory (SCNT), Institute of Modern Physics, Chinese Academy of Sciences, Huizhou 516000, Guangdong Province, China}}

\begin{document}
	
	
\title{An analysis of the gluon distribution with next-to-leading order splitting function in small-$x$}
	
\author{Jingxuan Chen}
\moe\iqm
\affiliation{Institute of Modern Physics, Chinese Academy of Sciences, Lanzhou 730000, China}
\affiliation{University of Chinese Academy of Sciences, Beijing 100049, China}

\author{Xiaopeng Wang}
\email{These authors contributed equally: Jingxuan Chen and Xiaopeng Wang.}
\affiliation{Institute of Modern Physics, Chinese Academy of Sciences, Lanzhou 730000, China}
\affiliation{Lanzhou University, Lanzhou 730000, China}
\affiliation{School of Nuclear Science and Technology, University of Chinese Academy of Sciences, Beijing 100049, China}

\author{Yanbing Cai}
\email{yanbingcai@mail.gufe.edu.cn}
\affiliation{Guizhou Key Laboratory in Physics and Related Areas, Guizhou University of Finance and Economics, Guiyang 550025, China}
\affiliation{Southern Center for Nuclear-Science Theory (SCNT), Institute of Modern Physics, Chinese Academy of Sciences, Huizhou 516000, China}
	
\author{Xurong Chen}
\email{xchen@impcas.ac.cn}
\affiliation{Institute of Modern Physics, Chinese Academy of Sciences, Lanzhou 730000, China}
\affiliation{School of Nuclear Science and Technology, University of Chinese Academy of Sciences, Beijing 100049, China}
	
\author{Qian Wang}
\email{qianwang@m.scnu.edu.cn}
\moe\iqm\scnt
	

\begin{abstract}
\newpage
An approximated solution for gluon distribution from DGLAP evolution equations with NLO splitting function in the small-$x$ limit is presented. We first obtain the simplified forms of LO and NLO splitting functions in the small-$x$ limit. With these approximated splitting functions, we obtain the analytical gluon distribution by using the Mellin transform. The free parameters in the boundary conditions are obtained by fitting the CJ15 gluon distribution data. We find that the asymptotic behavior of gluon distribution are consistent with the CJ15 data, however, the NLO results with the consideration of “ladder” structure of gluon emission are slightly better than those from LO. These results indicate that the corrections from NLO is significant and is necessary for a better description of the behavior of the gluon distribution in small-$x$ region. In addition, we investigate the DGLAP evolution of the proton structure function by using the analytical solution of the gluon distribution. The differential structure function shows that our results have a similar tendency with CJ15 at small-$x$.

\end{abstract}
	

\maketitle

\section{Introduction}
\label{sec:intro}
The deep inelastic scattering (DIS) is a very important process in high energy physics as it can define and measure the parton densities, which are instrumental to compute hard processes and test the quantum chromodynamics (QCD) \cite{Blumlein:2012bf}. Indeed, according to the collinear factorization theorem, the parton distribution functions (PDFs) are key ingredients to investigate the proton structure function and describe the cross sections at hard processes. In particular, the scattering cross sections can be expressed in terms of the parton densities, which are universal and process independent. An accurate determination of the PDFs is therefore necessary to improve our understanding of QCD. The PDFs are functions of the parton momentum fraction $x$ and the virtuality of the photon $Q^2$.  Generally, the parameterisation of PDFs is set at initial scale $Q_0^2$ and the value at a higher $Q^2$ is obtained through DGLAP evolution \cite{Altarelli:1977zs,Gribov:1972ri,Lipatov:1974qm,Dokshitzer:1977sg}. Therefore, the DGLAP evolution equations are standard tools for theoretical investigation of the PDFs.

The evolution equations for the PDFs are expressed in terms of the splitting functions that can be expanded in powers of the running coupling $\alpha_s$. By considering higher order of $\alpha_s$ , one can get a higher accuracy of PDFs. However, this does not mean that one can improve the accuracy without any limitation. On the one hand, deriving the higher order splitting functions is more tough. On the other hand, it is difficult to numerical and analytical solve the high order DGLAP evolution equation as they are complex integro-differential equations. In literature, the DGLAP equations can be numerically solved by iterating the evolution over infinitesimal steps in $x$ \cite{Tung:1988sn}, by using Mellin moments and their inversion \cite{Kosower:1997hg}, and by expanding parton distributions and splitting functions on Laguerre polynomials \cite{Schoeffel:1998tz}. There are various easy-to-use programs to perform the DGLAP evolution, e.g. the QCDNUM \cite{Botje:1999dj}, QCD-PEGASUS \cite{Vogt:2004ns}, CANDIA \cite{Cafarella:2005zj}. However, the numerical solutions with high accuracy are time-consumption. What's more, the numerical solution shall become worse at small-$x$ \cite{Coriano:1998wj,Devee:2012zz}. So, the analytical solution are more convenient to analyze the asymptotic behavior of the PDFs. To date, exact analytic solution of the DGLAP evolution equations in entire range of $x$ and $Q^2$ are difficult. However, one can solve them under certain conditions. Such as the gluon distributions have been obtained by using the Mellin transform under the double logarithmic approximation (DLA) \cite{Kovchegov:2012mbw}. The spin-independent and spin-dependent structure function are obtained by using Taylor expansion\cite{Baishya:2006iu,Baishya:2009zz}. The leading order singlet DGLAP evolution equations have been solved by using Laplace transform \cite{Block:2010fk,Block:2011xb}. These approximated solutions have successfully described the structure function under certain conditions, which improve our understanding of the asymptotic behavior of the PDFs.

The splitting functions are inputs for DGLAP evolution equations. The power of $\alpha_s$ in the splitting functions usually denote the order of DGLAP evolution equations. The leading order splitting functions $P_{gq}$ and $P_{gg}$ are singular at small-$x$. Thus the gluon distribution runs faster than the quark distribution in small-$x$, that is, the gluons are dominant in small-$x$. The analytical solution of DGLAP evolution in DLA for gluon distribution show a faster increase in small-$x$, which leads to an increase of quark distribution and eventually leads to an increase in structure function \cite{Kovchegov:2012mbw}. Therefore, the analysis of the asymptotic behavior of the PDFs is important to describe the measurement, e.g. structure function. In this study, we investigate the gluon distribution in small-$x$ limit up to next leading order. By using the Mellin transform, we obtain the approximated solution for the gluon distribution. We find that our analytical solution is consistent with the gluon distribution behavior of the CJ15 in small-$x$ and large $Q^2$ region. 

The rest of the paper is organized as follows. In the next section, we present the gluon distribution from DGLAP evolution with NLO splitting function and give a detailed derivation of the analytical solution in small-$x$ limit using the Mellin transform. In Sec.\ref{sec:result}, the fitting procedure is presented and the fit results are discussed. Then we investigate the DGLAP evolution of the proton structure function in Sec.\ref{sec:structure_function}. The main results of our study are summarized in Sec.\ref{sec:summary}.

\section{The Methods }
\label{sec:Hamiltonian}
Ignoring the quark contribution to the gluon rich distribution function at small x, the coupled DGLAP equation is
\begin{equation}
\label{eq:nlo-dglap}
\frac{\partial g(x,Q^2)}{\partial ln\frac{Q^2}{\Lambda^2}}|_{\mathrm{DGLAP}}=\int_{x}^{1}P_{\mathrm{gg}}(z)g\left(\frac xz,Q^2\right)\frac{\mathrm{d}z}{z},
\end{equation}
where the $P_{gg}$ is the splitting function. If consider up to NLO terms, $P_{gg}$ can be expanded as powers of $\alpha_s(Q^2)$ \cite{Kovchegov:2012mbw}
\begin{equation}
\begin{aligned}
P_\mathrm{gg}(z)&=\frac{\alpha_s(Q^2)}{2\pi}P_\mathrm{gg}^0(z)+
\left(\frac{\alpha_s(Q^2)}{2\pi}\right)^2P_\mathrm{gg}^1(z).
\end{aligned}
\end{equation}
Here the LO splitting function is
\begin{equation}
\label{eq:pgg0}
{P_\mathrm{gg}}^0(z)=2N_c\left(\frac{1-z}z+\frac {z}{(1-z)_+}+z(1-z)\right)+\left(\frac{11}2-\frac{2N_f}3\right)\delta(1-z),
\end{equation}
and the NLO splitting function is
\begin{equation}
\label{eq:pgg1}
\begin{aligned}
P_{gg}^{1}=& \begin{aligned}C_FT_f\left\{-16+8z+\frac{20z^2}{3}+\frac{4}{3z}-(6+10z)\ln z-2(1+z)\ln^2z\right\}\end{aligned}  \\
&+N_cT_f\left\{2-2z+\frac{26}9\left(z^2-\frac1 z\right)-\frac43(1+z)\ln z-\frac{20}9p(z)\right\} \\
&+N_c^2\left\{\frac{27(1-z)}2+\frac{67}9\left(z^2-\frac1 z\right)-\left(\frac{25}3-\frac{11z}3+\frac{44z^2}3\right)\ln z\right. \\
&+4(1+z)\ln^2z+\left(\frac{67}9+\ln^2z-\frac{\pi^2}3\right)p(z) \\
&-\left.4\ln z\ln(1-z)p(z)+2p(-z)S_2(z)\right\} .
\end{aligned}
\end{equation}
At small fraction momentum, the $z$-integral may get extra enhancement from the small-$z$ region\cite{Phukan:2017lzp}. So the splitting function can be approximated as
\begin{equation}
    P_{gg}(z)|_{z\ll1}\approx 2N_c+\frac{\alpha_s(Q^2)}{2\pi}H,
\end{equation}\
where the H defined as $H = C_F T_f (\frac{4}{3})+N_c T_f (-\frac{46}{9})$.

According to above approximation, the DGLAP equation can be rewritten as
\begin{equation}
\label{eq:NLO_DGLAP}
\frac{\partial g(x,Q^2)}{\partial ln\frac{Q^2}{\Lambda^2}}|_{\mathrm{DGLAP}}=\frac{\alpha_s(Q^2)}{2\pi}\int_{x}^{1} \left(2N_c+\frac{\alpha_s(Q^2)}{2\pi}H\right)g\left(\frac x z,Q^2\right)\frac{\mathrm{d}z}{z^2}.
\end{equation}
To obtain the the analytical solution for Eq.(\ref{eq:NLO_DGLAP}),  we can rewrite the DGLAP equation moment space by using the Mellin transform
\begin{equation}
\label{eq:melin_NLO_DGLAP}
Q^2\frac{\partial g_\omega(Q^2)}{\partial Q^2}=\frac{\alpha_s(Q^2)}{2\pi}(2N_c+\frac{\alpha_s(Q^2)}{2\pi}H)\frac1\omega g_\omega(Q^2) .
\end{equation}
Here $G_\omega(Q^2)$ is the gluon distribution in Mellin space
\begin{equation}
\label{eq:melin_xg}
g_\omega(Q^2)=\exp\left\{\int_{Q_0^2}^{Q^2}\frac{dQ^{\prime2}}{Q^{\prime2}}\left(\frac{\alpha_s(Q^2)}{2\pi}(2N_c+\frac{\alpha_s(Q^2)}{2\pi}H) \right) \frac{1}{\omega}\right\}g_\omega(Q_0^2) ,
\end{equation}
where the $g_\omega(Q_0^2)$ is the initial gluon distribution in Mellin space.
Respectively, the form of the running coupling constant in LO and NLO is 
\begin{equation}
\label{eq:alphasLO}
\frac{\alpha_s^\mathrm{LO}}{2\pi}=\frac{2}{\beta_0t},
\end{equation}

\begin{equation}
    \label{eq:alphasNLO}
    \frac{\alpha_s^{\mathrm{NLO}}}{2\pi}=\frac{2}{\beta_0t}\biggl[1-\frac{\beta_1\mathrm{ln}t}{\beta_0^2t}\biggr],
\end{equation}
where $\beta_0=\frac13(33-2N_f)$, $\beta_1=102-\frac{38}3N_f$, and $t=\ln\frac{Q^2}{\Lambda^2}$.
Inverting the Mellin transform, one obtain
\begin{equation}
\label{eq:invmelin_xg}
G(x,Q^2)=\int_{a-i\infty}^{a+i\infty}\frac{d\omega}{2\pi i}\exp\left\{\omega\ln\frac1x+\int_{Q_0^2}^{Q^2}\frac{dQ^{\prime2}}{Q^{\prime2}}\left(\frac{\alpha_s(Q^2)}{2\pi}(2N_c+\frac{\alpha_s(Q^2)}{2\pi}H) \right)\frac{1}{\omega}\right\}g_\omega(Q_0^2),
\end{equation}
where $G(x,Q^2)=xg(x,Q^2)$.
Here, we define
\begin{equation}
\label{eq:rho}
\rho(Q^2)=\int_{Q_0^2}^{Q^2}\frac{dQ^{\prime2}}{Q^{\prime2}}\left(\frac{\alpha_s(Q^{\prime2})}{2\pi}(2N_c+\frac{\alpha_s(Q^{\prime2})}{2\pi}H) \right),
\end{equation}

\begin{equation}
\label{eq:Pw}
P(\omega)=\omega\ln\frac1x + \frac{\rho(Q^2)}{\omega}.
\end{equation}
To analyze the asymptotic behavior of the gluon distribution in Eq.(\ref{eq:invmelin_xg}) in small fraction momentum limit, it is needed to find the saddle points of the exponent $P(\omega)$, which are defined by the condition
$P^{\prime}(\omega=\omega_{sp})=0,$
one obtain the saddle points
\begin{equation}
\label{eq:omega_sp}
\omega_{sp}=\pm \sqrt{\frac{\rho(Q^2)}{\ln(1/x)}}.
\end{equation}
Making a Taylor expansion around the saddle point, the Eq.(\ref{eq:Pw}) would be
\begin{equation}
\label{eq:p_tale}
P(\omega)\approx P(\omega_{sp})+\frac12P^{\prime\prime}(\omega_{sp})(\omega-\omega_{sp})^2 ,
\end{equation}
where the term $\omega-\omega_{sp}$ is zero at the saddle points. 
A new integration variable $\omega$ is defined by $\omega-\omega_{sp}\equiv iw$\cite{Kovchegov:2012mbw} and the Eq.(\ref{eq:invmelin_xg}) becomes 
\begin{equation}
\label{eq:invmeilin_xg_2_order}
G(x,Q^2)\approx e^{P(\omega_{sp})}g_{\omega_{sp}}(Q_0^2)\int_{-\infty}^\infty\frac{dw}{2\pi}e^{-P^{\prime\prime}(\omega_{sp})w^2/2}.
\end{equation}
The result of performing $\omega$-integration is
\begin{equation}
G(x,Q^2)\approx \frac{g_{\omega_{sp}}(Q_0^2)}{\sqrt{4\pi}}(\rho)^{1/4}(\ln(1/x))^{-3/4} \times \textrm{exp}\left(2\sqrt{\rho\ln(1/x)} \right).
\end{equation}
For small-$x$ and large $Q^2$, the gluon distribution could be rewritten as
\begin{equation}
\label{eq:G_exp}
G(x,Q^2)\sim \,\textrm{exp}\left(2\sqrt{\rho\ln(1/x)}\right),
\end{equation}
where it is a solution of NLO or LO of the DGLAP equation depends on $\rho(Q^2)$.
in LO,  $\rho(Q^2)$ is expressed as
\begin{equation}
    \label{eq:lo_rho}
    \rho^{\mathrm{LO}}(Q^2)=\frac{4N_c}{\beta_0}\ln \frac{t}{t_0},
\end{equation}
where $t=\ln\frac{Q^2}{\Lambda^2}$ and $t_0=\ln \frac{Q_0^2}{\Lambda^2}$. Although it is diffcult to obtain an analytic expression for $\rho(Q^2)$ in NLO directly, the relation between $\rho^{\mathrm{NLO}}(Q^2)$ and $\rho^{\mathrm{LO}}(Q^2)$ can be expressed as, 
\begin{equation}
    \label{eq:ratio_rho_nlo_lo}
    \frac{\rho^{\mathrm{NLO}}(Q^2)}{\rho^{\mathrm{LO}}(Q^2)}=\mathrm{R}(Q^2).
\end{equation}

From Ralston's calculations.~\cite{Ralston:1986hr}, a boundary conditions ($K(Q^2)$) is provided, gluon distribution is written as
\begin{equation}
\label{eq:G_K_exp}
    \begin{aligned}
    & G(x,Q^2)=K(Q^2)\times\textrm{exp}\left(2(\rho\ln(1/x))^{1/2}\right), \\
    & K\left(Q^2\right)=a\left[\exp \left(\xi-\xi_0\right)+ b\right]\times \exp \left[c\left(\xi-\xi_0\right)^{1 / 2}\right] ,
    \end{aligned}
\end{equation}
where $\xi=\ln \ln (Q^2/\Lambda^2_{QCD})$ and $\xi_0=\ln \ln (Q_0^2/\Lambda^2_{QCD})$. In addition, when gluon emission kernel have a ``ladder" structure, gluon distribution can be considered as \cite{kovchegov_levin_2012}
\begin{equation}
\label{eq:G_I0}
G(x,Q^2)\sim \,\mathrm{I_0}\left(2\sqrt{\rho\ln(1/x)}\right),
\end{equation}
where $\mathrm{I}_0$ is modified Bessel function. Therefore, gluon distribution in Eq.\,(\ref{eq:G_K_exp}) is rewritten as
\begin{equation}
\label{eq:G_K_I0}
    \begin{aligned}
    & G(x,Q^2)=K(Q^2)\times \mathrm{I_0}\left(2(\rho\ln(1/x))^{1/2}\right).
    \end{aligned}
\end{equation}
In the next section, we shall use our analytical results to fit the gluon distribution data. In our fits, we allow the value of the saddle point to shift slightly. So, we replace $\mathrm{I_0}\left(2(\rho\ln(1/x))^{1/2}\right)$  with $\mathrm{I_0}\left(2(\rho\ln(1/x))^{d}\right)$ and $\mathrm{exp}\left(2(\rho\ln(1/x))^{1/2}\right)$ with $\mathrm{exp}\left(2(\rho\ln(1/x))^{d}\right)$ in Eq.\,(\ref{eq:G_K_I0}) and Eq.\,(\ref{eq:G_K_exp}), where $d$ is a free parameter float around 0.5.

\section{Fitting results}
\label{sec:result}

\begin{figure*}[htbp]
    \centering
    \subfigure[]
    {
        \begin{minipage}[htbp]{.4\linewidth}
	\centering
	\includegraphics[scale=0.8]{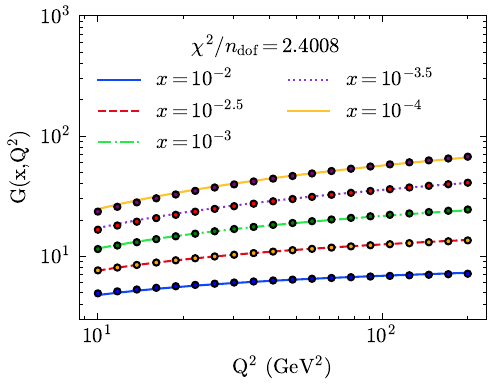}
	\end{minipage}
    }
    \subfigure[]
    {
        \begin{minipage}[htbp]{.4\linewidth}
	\centering
	\includegraphics[scale=0.8]{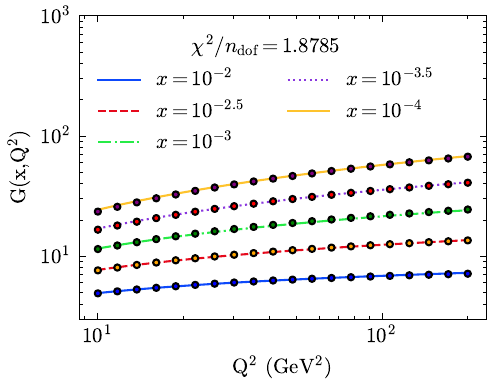}
	\end{minipage}
    }
	\caption{Fitting to the CJ15lo gluon distribution with 4 parameters in $10\,\mathrm{GeV^2}<Q^2<200\,\mathrm{GeV^2}$ and $10^{-4}<x<10^{-2}$ with Eq.\,(\ref{eq:G_K_exp}) in (a) and with Eq.\,(\ref{eq:G_K_I0}) in (b) in LO.}
    \label{fig:ol}
\end{figure*}

In this study, we use the LO of the fitting equation to fit the CJ15LO data \cite{Buckley:2014ana,Accardi:2016qay} in $10\,\mathrm{GeV^2}<Q^2<200\,\mathrm{GeV^2}$ and $10^{-4}<x<10^{-2}$  region. The fitting results are shown in Table.\,\ref{tab:table1} where S1 (S2) stands for the fitting parameters of Eq.\,(\ref{eq:G_K_exp}) (Eq.\,(\ref{eq:G_K_I0})) when $Q^2_0=1\,\mathrm{GeV^2}$. From the (a) of the Fig.~\ref{fig:ol}, the fitting result with Eq.\,(\ref{eq:G_K_exp}) is presented. As shown in the (b) of the Fig.~\ref{fig:ol}, our modified Ralston's solution in the situation of S2 is agreement well with CJ15LO gluon distribution. On the other hand, the value of d does not deviate by 0.5 too much.

\begin{table}[htbp]
	\caption{\label{tab:table1}%
		Parameters from the fit to the  CJ15LO and CJ15NLO \cite{Accardi:2016qay}  in LO and NLO  respectively are shown. S1 represents the fitting parameters of Eq.\,(\ref{eq:G_K_exp}) and S2 stands for the fitting parameters of Eq.\,(\ref{eq:G_K_I0}).
	}
		\begin{ruledtabular}
		\begin{tabular}{ccccc}
			&LO(S1)&LO(S2)&NLO(S1)&NLO(S2) \\
			\colrule
			\textrm{$a$} & $52.579\,\pm \,3.442$ & $185.446\,\pm \,3.254$ &20.307 $\pm \,1.899$ &108.199 $\pm \,8.763$\\	
			\textrm{$b$}& $-1.205\,\pm \,0.014$ & $-1.174\,\pm \,0.004$ &-0.978 $\pm \,0.017$ &-1.023 $\pm \,0.013$\\
			\textrm{$c$}& $-7.639\,\pm \,0.058$ & $-7.463\,\pm \,0.016$ &-6.963 $\pm \,0.095$ &-7.186 $\pm \,0.084$\\
			\multicolumn{1}{c}{$d$}& $0.505\,\pm \,2\times10^{-4}$ & $0.522\,\pm \,2\times10^{-4}$ &0.428 $\pm \,10^{-3}$ &0.452 $\pm \,10^{-3}$\\
			$\chi^2/dof$ & $2.4008$ &$1.8785$& 4.411&1.714\\
		\end{tabular}
		\end{ruledtabular}
\end{table}

\begin{figure*}[htbp]
    \centering
    \subfigure[]
    {
        \begin{minipage}[htbp]{.4\linewidth}
	\centering
	\includegraphics[scale=0.8]{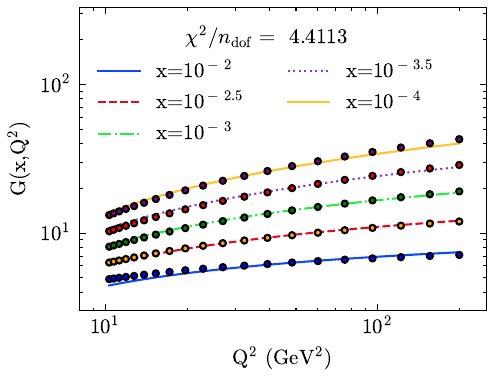}
	\end{minipage}
    }
    \subfigure[]
    {
        \begin{minipage}[htbp]{.4\linewidth}
	\centering
	\includegraphics[scale=0.8]{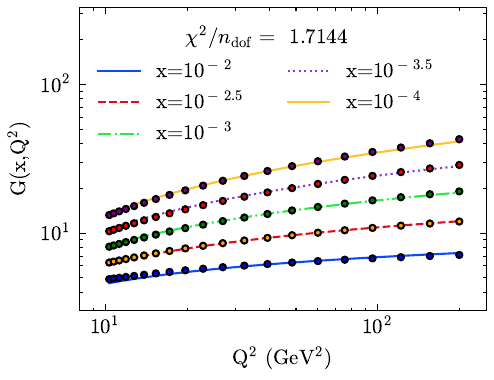}
	\end{minipage}
    }
	\caption{Fitting to the CJ15nlo gluon distribution with 4 parameters in $10\,\mathrm{GeV^2}<Q^2<200\,\mathrm{GeV^2}$ and $10^{-4}<x<10^{-2}$ with Eq.\,(\ref{eq:G_K_exp}) in (a) and with Eq.\,(\ref{eq:G_K_I0}) in (b) in NLO.}
   \label{fig:fit_NLO_CJ15nlo_exp}
\end{figure*}

When NLO correction is considered, $\rho^{\mathrm{NLO}}$ is used. As described in the previous section, it is difficult to get an analytic form of $\rho^{\mathrm{NLO}}$. But it can be described with Eq.\,(\ref{eq:ratio_rho_nlo_lo}) and Eq.\,(\ref{eq:lo_rho}) as 
\begin{equation}
    \label{eq:NLO_lo_represent}
    \rho^{\mathrm{NLO}}=\left(\frac{4N_c}{\beta_0}\ln \frac{t}{t_0}\right)R(t).
\end{equation} 
We find that $R(t)$ can be written as $\frac{\tilde{a}t^{\tilde{d}}}{\tilde{b}+\tilde{c}t^{\tilde{e}}}$ in our fitting region.
When $Q^2_0=2.5\,\mathrm{GeV}^2$, $\tilde{a}=13.115,\,\tilde{b}=14.638,\,\tilde{c}=19.065,\tilde{d}=0.665 \,\mathrm{and}\,\tilde{e}=0.616$. Then one can obtain a adequate fitting results  with 4 parameters in the Fig.~\ref{fig:fit_NLO_CJ15nlo_exp} in $10\,\mathrm{GeV^2}<Q^2<200\,\mathrm{GeV^2}$ and $10^{-4}<x<10^{-2}$ with Eq.\,(\ref{eq:G_K_exp}) and with Eq.\,(\ref{eq:G_K_I0}) respectively. The values of the fitting parameters are also listed in Table 1. The data of gluon distribution comes from CJ15nlo. in larger $Q^2$ region, our models are in better agreement with CJ15nlo. And compared with the fitting result with Eq.\,(\ref{eq:G_K_exp}), the gluon distribution from Eq.\,(\ref{eq:G_K_I0}) is more suitable. And it notes that the value of d is less than 0.5 but not not far away. Overall, our models are a valid description of CJ15nlo in the certain region.

\section{ The differential structure function from gluon distribution}
\label{sec:structure_function}
 In small-x region, when the contribution of gluon is only considered, the DGLAP evolution equation for proton structure function $\rm F_2$ can be written as \cite{Ball:1994du,Boroun:2013xwt,Boroun:2006xg}
 \begin{equation}\begin{aligned}
\frac{\partial F_{2}(x,Q^{2})}{\partial\ln Q^{2}} =
\frac{20}{9}\frac{\alpha_{s}}{2\pi}\int_{x}^{1}dzP_{qg}(z)G\left(\frac{x}{z},Q^{2}\right).
\end{aligned}\end{equation}
Here, the splitting function $P_{qg}(z)$ is defined as
\begin{equation}
\label{eq:pqg}
    P_{qg}(z)=P^{\mathrm{LO}}_{qg}(z)+\frac{\alpha_s}{2\pi}P^{\mathrm{NLO}}_{qg}(z),
\end{equation}
where splitting functions can be found in \cite{Boroun:2013xwt}.
When $P^{\mathrm{LO}}_{qg}(z)$ is only contained, by expanding gluon distribution around $z=\frac{1}{2}$, Prytz's approach for $\rm F_2$ evolution function is written as \cite{Prytz:1993vr}
\begin{equation}
    \label{eq:prytzlo}
    \frac{\partial F_{2}(x,Q^{2})}{\partial\ln Q^{2}}|_{\mathrm{LO}} \approx \frac{20}{27}\frac{\alpha_s}{2\pi}G(2x).
\end{equation}
When based on expanding the gluon distribution around $z=0$, a adequate Bora et al's approach is given as \cite{Bora:1995np,GayDucati:1996ry}
\begin{equation}
    \label{eq:Bora}
    \frac{\partial F_{2}(x,Q^{2})}{\partial\ln Q^{2}}|_{\mathrm{LO}} \approx \frac{20}{27}\frac{\alpha_s}{2\pi}G(\frac{3}{2}x).
\end{equation}

For the consideration of the next leading order correction, $\frac{\partial F_{2}(x,Q^{2})}{\partial\ln Q^{2}}$ is given as
 \begin{equation}
 \label{eq:F2_lo_nlo_one}
 \begin{aligned}
\frac{\partial F_{2}(x,Q^{2})}{\partial\ln Q^{2}} \approx
\frac{\partial F_{2}(x,Q^{2})}{\partial\ln Q^{2}}|_{\mathrm{LO}}+
\frac{20}{9}\left(\frac{\alpha_{s}}{2\pi}\right)^2\int_{x}^{1}dzP^{\mathrm{NLO}}_{qg}(z)G\left(\frac{x}{z},Q^{2}\right).
\end{aligned}\end{equation}

A simple form of $p^{\mathrm{NLO}}$ could be considered in small-$x$ limit \cite{Boroun:2006xg,ellis_stirling_webber_1996}, 
\begin{equation}
    \label{eq:simple_nlo}    p^{\mathrm{NLO}}\rightarrow\frac{\alpha_s}{2\pi}\frac{60}{9z}.
\end{equation}
Therefore, $\frac{\partial F_{2}(x,Q^{2})}{\partial\ln Q^{2}}$ is rewritten as
 \begin{equation}
 \label{eq:F2_lo_nlo}
 \begin{aligned}
\frac{\partial F_{2}(x,Q^{2})}{\partial\ln Q^{2}} \approx
\frac{\partial F_{2}(x,Q^{2})}{\partial\ln Q^{2}}|_{\mathrm{LO}}+
\frac{1200}{81} \left(\frac{\alpha_{s}}{2\pi}\right)^2 \int_{x}^{1}dz \frac{1}{z}G\left(\frac{x}{z},Q^{2}\right).
\end{aligned}\end{equation}
Here, we expand the second part of the right-hand side of the Eq.\,(\ref{eq:F2_lo_nlo}) around any point $\alpha$ in the same way \cite{GayDucati:1996ry}.
$\int_{x}^{1}dz \frac{1}{z}G\left(\frac{x}{z},Q^{2}\right)$ can be written as
\begin{equation}
    \label{eq:expand_NLO}
    \begin{aligned}
        &\int_{x}^{1}dz \frac{1}{z}G\left(\frac{x}{z},Q^{2}\right)
        &=\frac{1}{x}\int_{x}^{1}dz \tilde{G}\left(\frac{x}{z},Q^{2}\right)=\frac{1}{x}\int^{1-x}_{0}d\tilde{z} \tilde{G}\left(\frac{x}{1-\tilde{z}},Q^{2}\right),
    \end{aligned}
\end{equation}
where $\tilde{G}(x,Q^2)=xG(x,Q^2)$. Then, $\tilde{G}$ is expanded to the first order at $z=\alpha$, we get 
\begin{equation}
\label{eq:expand_NLO_1}
    \begin{aligned}
        \frac{1}{x}\int^{1-x}_{0}d\tilde{z} \tilde{G}\left(\frac{x}{1-\tilde{z}},Q^{2}\right)\approx \frac{1}{x}\int^{1-x}_{0}d\tilde{z} \left(\tilde{G}\left(\frac{x}{1-\alpha},Q^{2}\right)+\frac{x(z-\alpha)}{(\alpha-1)^2}\tilde{G}\left(\frac{x}{1-\alpha},Q^{2}\right)\right).
    \end{aligned}
\end{equation}
In small-x limit, Eq.\,(\ref{eq:expand_NLO_1}) is written as
\begin{equation}
   \label{eq:expand_NLO_2}
   \frac{1}{x}\int^{1-x}_{0}d\tilde{z} \tilde{G}\left(\frac{x}{1-\tilde{z}},Q^{2}\right)\approx \frac{\left(1-x\right)}{x} \tilde{G}\left(\frac{\frac{3}{2}-2\alpha}{(1-\alpha)^2}x\right)=\left(1-x\right) \frac{\frac{3}{2}-2\alpha}{(1-\alpha)^2}G\left(\frac{\frac{3}{2}-2\alpha}{(1-\alpha)^2}x\right).
\end{equation}

\begin{figure*}[htbp]
    \centering
    \includegraphics[scale=1.5]{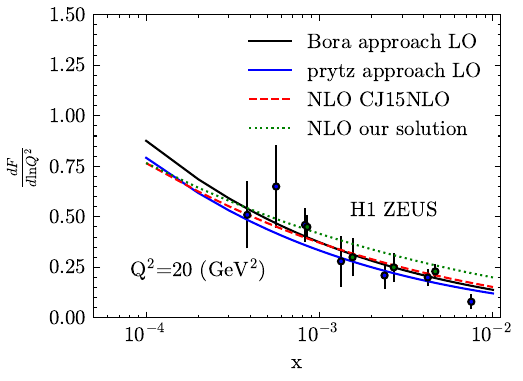}
    \caption{The differential structure function. Experimental data from H1 and ZEUS. Black and blue solid line stands for the calculation of differential structure function with Bora's approach and Prytz's approach in LO respectively. Red dashed line represents the calculation of differential structure function using CJ15NLO data \cite{Accardi:2016qay} directly with Eq.\,(\ref{eq:F2_lo_nlo_one}). Green Dotted line means that our approximated solution is used with Eq.(\ref{approx_NLO})}
        \label{fig:dfdQ2}
\end{figure*}

Therefore, from Eq.\,(\ref{eq:F2_lo_nlo}) we have
\begin{equation}
    \label{approx_NLO}
    \frac{\partial F_{2}(x,Q^{2})}{\partial\ln Q^{2}} \approx
\frac{20}{27}\frac{\alpha_s}{2\pi}G(2x)+\frac{1200}{81} \left(\frac{\alpha_{s}}{2\pi}\right)^2 \left(1-x\right) \frac{\frac{3}{2}-2\alpha}{(1-\alpha)^2}G\left(\frac{\frac{3}{2}-2\alpha}{(1-\alpha)^2}x\right).
\end{equation}
As shown in Fig.\,\ref{fig:dfdQ2}, the calculations of $\frac{\partial F_{2}(x,Q^{2})}{\partial\ln Q^{2}}$ are presented in LO and NLO. Black and blue solid line stands for the calculation of differential structure function with Bora's approach and Prytz's approach in LO respectively. Red dashed line represents the calculation of differential structure function using CJ15NLO data \cite{Accardi:2016qay} directly with Eq.\,(\ref{eq:F2_lo_nlo_one}) where no approximation is made to NLO splitting function. Green Dotted line means that our approximated solution is used with Eq.(\ref{approx_NLO}) when $\alpha=0.0$. We find that the result is better when $\alpha=0.0$. It notes that Prytz's approach with NLO running coupling constant is used for $\frac{\partial F_{2}(x,Q^{2})}{\partial\ln Q^{2}}|_{\mathrm{LO}}$ when NLO is considered. For LO results, the Fig.\,\ref{fig:dfdQ2} shows that Prytz's approach and Bora's approach are valid.  And when NLO correction is contained, our result with  approximated solution is higer than the calculation of differential structure function using CJ15NLO data directly with Eq.\,(\ref{eq:F2_lo_nlo_one}). When $x$ decreases, the gap between these two calculations become narrowing. 

\section{Summary}
\label{sec:summary}
In this study, we use LO and NLO approximated solutions from the DGLAP equation to fit the gluon distribution of CJ15LO and CJ15NLO respectively. The fitting results presents that the LO and NLO approximated solutions are satisfactory in small-x region and large $Q^2$ region. Compared with the fitting
result with Eq.(21), the gluon distribution from Eq.(23) in which ``ladder'' structure of gluon emission is considered is better. 
Then, $\frac{\partial F_{2}(x,Q^{2})}{\partial\ln Q^{2}}$ is calculated with the LO and NLO approximated solutions at $Q^2=20\,\mathrm{GeV^2}$. On the other hand, $\frac{\partial F_{2}(x,Q^{2})}{\partial\ln Q^{2}}$ is also given by Eq.\,(\ref{eq:F2_lo_nlo}) with CJ15NLO gluon distribution directly. Our result from approximated expression Eq.\,(\ref{approx_NLO}) is higer than the calculation from CJ15NLO gluon distribution in larger $x$. This suggests that the approximation in large $x$ is not valid. However, our approximated results have a similar tendency with CJ15 at small-$x$, which indicates the approximation is valid to describe the asymptotic behavior in small-$x$.

\begin{acknowledgments}
This work is supported by the Strategic Priority Research Program of Chinese Academy of Sciences under Grant No. XDB34030301; National Natural Science Foundation of China under Grant No. 12375073; Major Project of Basic and Applied Basic Research in Guangdong Province under Grant No. 2020B0301030008; Guizhou Provincial Basic Research Program (Natural Science) under Grant No. QKHJC-ZK[2023]YB027; Education Department of Guizhou Province under Grant No. QJJ[2022]016.
\end{acknowledgments}

\bibliographystyle{apsrev4-1}
\bibliography{ref}

\end{document}